\documentclass[aps,pre,twocolumn,groupedaddress]{revtex4-1}
\pdfoutput=1
\usepackage[utf8]{inputenc}
\usepackage{graphicx}
\usepackage{amsmath}
\usepackage{color}
\usepackage{array}
\usepackage{color}
\usepackage{amsfonts}
\usepackage{amssymb}
\usepackage{mathtools} 
\usepackage{breakurl}

\begin{document}


\title{Multifractality in the fidelity of the Toffoli gate}

\author{Jalil Khatibi Moqadam}
\author{Guilherme S. Welter}
\author{Paulo A. A. Esquef}
\affiliation{Laborat\'orio Nacional de Computa\c{c}\~{a}o Cient\'ifica (LNCC),
             Petrópolis, RJ, Brazil}


\date{\today}

\begin{abstract}

We analyze the multifractality of the fidelity in an engineered Toffoli gate.
Using quantum control methods, we define several optimization problems whose global solutions
realize the gate in a chain of three qubits with $XY$ Heisenberg interaction. We perturb the
system by introducing imperfections in the form of $1/f$ noise to the interqubit couplings.
Multifractal analysis shows that the degree of multifractality in the gate fidelity increases
when optimized control pulses used to render the fidelity are less sensitive to variations in
the interqubits coupling strengths.

\end{abstract}


\maketitle

\section{Introduction}
The multifracal formalism describing the scaling of the moments for some distributions in complex
systems has been widely used in studying a variety of classical systems \cite{mandelbrot1974intermittent,
halsey1986fractal,meneveau1991multifractal,muzy1991wavelets,mandelbrot1997multifractal,
stanley1988multifractal,abry2002multiscale,jafari2007long,lovejoy2012haar,ausloos2012generalized}.
Recently, multifractality has also appreciated in quantum systems.
Quantum wave functions in the Anderson model show multifractality at metal-insulator
transition \cite{mirlin2000statistics,
evers2008anderson,rodriguez2009Multifractal,rodriguez2009optimisation,
rodriguez2010critical,rodriguez2011multifractal,burmistrov2013multifractality}.
Wave functions in the quantum Hall transition are also multifractal \cite{huckestein1995scaling,
evers2001multifractality,evers2003multifractality}.
Actually, the strong fluctuations of the wave function amplitude are characterized as
wave function multifractality. The relevant normalized measure is the squared modulus of the wave
function $|\psi(\mathbf{r})|^2$ and the corresponding moments
are $p_q = \int d \mathbf{r} |\psi(\mathbf{r})|^{2q}$, the so-called inverse participation ratios.

Other examples of the multifractal wave functions are certain eigenstates of the quantum baker's
map~\cite{meenakshisundaram2005multifractal}, the eigenfunctions of one dimensional intermediate
quantum maps~\cite{martin2008multifractality}, the eigenfunction of Anderson
map \cite{martin2010multifractality},
the Floquet spectrum~\cite{Bandyopadhyay2010Generating}, the electronic states in the Fibonacci
superlattice under weak electric fields~\cite{woloszyn2012multifractal} and the individual wave
packets in a periodically kicked system \cite{garcia2012multifractality}.
Moreover, an ensemble of random matrices can be constructed such that the corresponding
eigenvectors become multifractal \cite{mirlin1996transition,kravtsov1997new,
bogomolny2012multifractal,fyodorov2009anderson}.

Other measures have also been applied to characterize the
multifractality in quantum systems. The Rényi entropy was used to study the
multifractality in the ground state wave function in the spin
chains~\cite{atlas2012multifractality,luitz2014universal}.
The von Neumann entanglement entropy also used to analyze the multifractality in
the wave functions at localization transition~\cite{jia2008multifractal}
and also in the entanglement of random states~\cite{giraud2009entropy}.
The quantum fidelity is another measure that has already been used to analyze the
fractal properties in periodically kicked quantum
systems~\cite{pellegrini2007fractal,bin2010fractals}. The quantum fidelity is defined
as the overlap between the perturbed and the unperturbed quantum
states $| \langle \psi_{\epsilon}(t) | \psi(t) \rangle|$.

In this paper, we introduce the gate
fidelity $\left| Tr \left[ U_{\epsilon}^{\dagger}(t) U(t) \right] \right|$
to study the multifractality in quantum gates. Specifically the Toffoli gate,
a three-qubit gate with central role in quantum information processing is
considered here. The gate is realized by applying a sequence of optimized control
pulses in a system of three interacting qubits~\cite{stojanovic2012quantum}.
We perturb the system by adding $1/f$ noise to the interqubit couplings
and then implement the gate for a large number of noise realizations.
Such a noise model has already been discussed in~\cite{moqadam2013analyzing}.
The resulted fidelity sequence is then analyzed numerically 
in the multifractal framework, using the formalism recently proposed
in~\cite{welter2013multifractal}.

By manipulating the objective functional in the quantum optimization problem we design several
new gates which show higher degree of multifractality compared with the gate originally proposed 
in Ref.~\cite{stojanovic2012quantum}. 
More specifically, it is shown that, by decreasing the sensitivity of the gate
fidelity with respect to variations in the interqubits coupling strengths, the complexity the
system increases and, as a consequence of it, the degree of multifractality in the gate fidelity
also increases.

The paper is organized as follows. In Sec.~\ref{sec:toffoli} five different realizations of the
Toffoli gate are characterized. The multifractal formalism is introduced in Sec.~\ref{sec:multifractal}.
In Sec.~\ref{sec:mf_fidelity} the numerical analysis of the multifractality in the fidelity
of the Toffoli gate is reported. Finally, the summary and discussions are presented in
Sec.~\ref{sec:conclusion}.

\section{\label{sec:toffoli}The Toffoli Gate}
The Toffoli gate is an element of the special unitary group $SU(8)$ equal to the identity
matrix $\mathcal{I}_{8 \times 8}$ except for the last two rows which are interchanged.
It affects three-qubit states belonging to the eight-dimensional Hilbert
space $\mathbb{C}^8$. The Toffoli gate can be implemented in a system of coupled qubits
using different methods. We consider a system of three mutually coupled qubits and apply
a sequence of optimized pulses that affect all the individual qubits.
Suppose the chain of interacting qubits is described by a Heisenberg $XY$ Hamiltonian
\begin{equation}
 \label{eq:XY-type}
 H_0 = \sum_{m<l}J_{ml}\left({\sigma}_{mx}{\sigma}_{lx}+{\sigma}_{my}{\sigma}_{ly}\right),
 \;\;\;\;\;m,l=1,2,3
\end{equation}
where $J_{ml}$ are the interqubit coupling strength and ${\sigma}_{mx}$ and ${\sigma}_{my}$
are Pauli $X$ and $Y$ matrices for qubit $m$.

The chain of qubits can be manipulated by the control Hamiltonian
\begin{equation}
 \label{eq:zeeman}
 H_c(t)=\sum_{m=1}^{3}[{u^{(m)}_{x}(t)} {\sigma_{mx}}+{u^{(m)}_{y}(t)}{\sigma_{my}}],
\end{equation}
where $u^{(m)}_{x}(t)$ and $u^{(m)}_{y}(t)$ are two different types of control fields
affecting the individual qubits.

The system dynamics is therefore governed by the sum of Hamiltonians in
Eqs.~(\ref{eq:XY-type}) and (\ref{eq:zeeman}).
The Schr\"odinger equation for the unitary operators ($\hbar=1$)
\begin{equation}
\label{eq:Schrodinger_operator_eq}
\left\{
\begin{array}{ll}
 dU/dt = -i \left( H_0 + H_c \right) U  \\
 U(0) = \mathcal{I}_{8 \times 8},
\end{array}
 \right.
\end{equation}
is used to obtain the evolution operator of the system.

Specifying the control fields such that the evolution operator in a given time interval $t=t_g$
implements the Toffoli gate is a numerical optimization problem.
Here, the control fields are considered piecewise constant functions of time and the
gate time is divided into $N_t$ equal pieces accordingly. The Schr\"odinger equation
can then be solved straightforwardly in each time interval. The total time evolution
operator is obtained by multiplying the partial time evolution operators in the reverse
order. The fidelity is defined as
\begin{equation}
\label{eq:fidelity}
F=\frac{1}{8}\left|\;\mathrm{Tr}\left[U^\dagger\left(t_g,N_t,\mathbf{u},\{J_{ml}\}\right)
U_{\mathrm{Toff}} \right]\;\right|,
\end{equation}
where $U$ is the the total time evolution of the system during $t=t_g$, $\mathbf{u}$ is the
concatenation of all control pulses and $U_{\mathrm{Toff}}$ is the Toffoli gate.
The values of the control pulses are obtained by solving the optimization problem
\begin{equation}
 \label{eq:optimization_original}
 \max_{ \mathbf{u} }{ F\left( \mathbf{u} \right) }.
\end{equation}

We obtain five sets of control pulses using fidelity~(\ref{eq:fidelity}) in different optimization
problems. The control pulses in each set are optimized such that the resulting gate fidelity
functional has a specific response to variations in the interqubit couplings $J_{ml}$.
Each set corresponds to a different realization of the Toffoli gate.

The first set of control pulses, $\mathbf{u}_1$,
is the global solution of problem~(\ref{eq:optimization_original}) with $N_t=20$,
$J_{12}=J_{23}=6J_{13}=\bar{J}$ and $t_g = 4.18 \bar{J}^{-1}$.
Finding such a set of control pulses has been fully addressed in Ref.~\cite{stojanovic2012quantum}.
The set is composed of 60 control pulses implementing the Toffoli gate with a fidelity above $99\%$.

\begin{figure}
\includegraphics[trim = 13mm 68mm 20mm 65mm, clip=true, width=9cm]{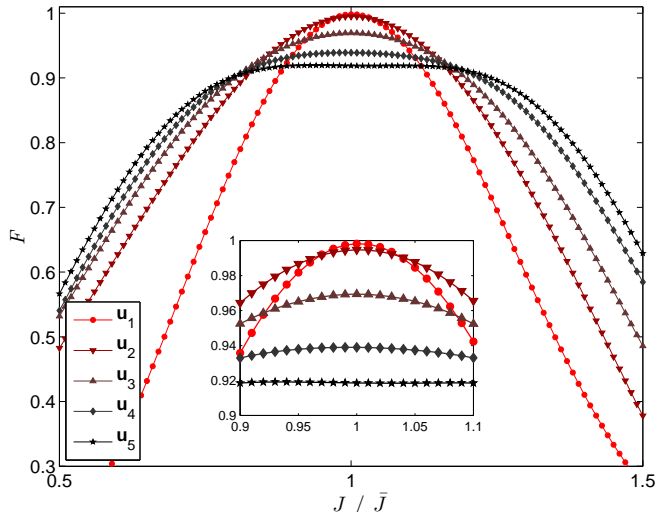}
\caption{\label{fig:fid_ver_j}(Color online) The fidelity versus $J/\bar{J}$ for five different sets
of control pulses (see the text for the definitions of $\mathbf{u}_1$ to $\mathbf{u}_5$).
Inset: a zoom on the region $|J/\bar{J}-1|\leq 0.1$.}
\end{figure}

The curve marked with $\bullet$ in Fig.~(\ref{fig:fid_ver_j}) depicts the
\nobreak fidelity~[Eq.~(\ref{eq:fidelity})]
in terms of $J/\bar{J}$ for the set $\mathbf{u}_1$ whose elements are optimized for $J=\bar{J}$.
The fidelity is clearly high in the vicinity of $J=\bar{J}$ and decreases quickly by
deviating from that point.

The second set of control pulses, $\mathbf{u}_2$, is the global solution of the problem
\begin{equation}
 \label{eq:optimization_interval}
 \max_{ \mathbf{u} }{ \int_{\bar{J}-\delta J}^{\bar{J}+\delta J}F(\mathbf{u},J)w(J)\,dJ},
\end{equation}
where
\begin{equation}
 \label{eq:weight_function}
 w(J)= \left\{ \begin{array}{ll}
              0, \;\;\;\;\; & \left| \frac{J}{\bar{J}}-1 \right| \leq \delta_1\\
              \;\\
              1, \;\;\;\;\; \delta_1  < & \left| \frac{J}{\bar{J}}-1 \right| \leq \delta_2,
              \end{array} \right.
\end{equation}
with $\delta_1=0.05$, $\delta_2 = 0.15$ and $\delta J = 0.15 \bar{J}$.
The detailed discussion for optimization problem~(\ref{eq:optimization_interval}) has been given
in Ref.~\cite{moqadam2013analyzing}.

The curve marked with {\small $\blacktriangledown$} in Fig.~(\ref{fig:fid_ver_j}) depicts the fidelity
in terms of $J/\bar{J}$ for the set $\mathbf{u}_2$.  In this case, compared with the
fidelity diagram for the set $\mathbf{u}_1$, the fidelities are smaller in the vicinity
of $J=\bar{J}$ but larger in other points. Applying the pulses in the set $\mathbf{u}_2$ leads to a
Toffoli gate which is less sensitive to the variation in $J$, specially when $|J/\bar{J}-1|\leq 0.1$.

The last three sets of control pulses are obtained by
finding the global solutions of the optimization problem
\begin{align}
 \label{eq:optimization_flattening}
 \max_{\mathbf{u}}
        \biggl \{ &\beta \left[ F(\mathbf{u},\bar{J}-J_0) +
                                 F(\mathbf{u},\bar{J}) +
                                 F(\mathbf{u},\bar{J}+J_0) \right]   \nonumber  \\
                  &\hspace{-0.1cm}-  \left| 2F(\mathbf{u},\bar{J}) -
                                                  F(\mathbf{u},\bar{J}-J_0) -
                                                  F(\mathbf{u},\bar{J}+J_0) \right| \nonumber  \\
                  &\hspace{-0.1cm}-
                                     \left| F(\mathbf{u},\bar{J}-J_0) -
                                            F(\mathbf{u},\bar{J}+J_0) \; \right| \biggr \},
\end{align}
where $J_0=0.1\bar{J}$. Setting $\beta \approx 10^3,10,0.1$ gives the sets $\mathbf{u}_3$,
$\mathbf{u}_4$ and $\mathbf{u}_5$ respectively.
The main interest in optimization problem~(\ref{eq:optimization_flattening}) is to find those
solutions whose fidelity functionals are almost flat in the interval $\left[\bar{J}-J_0,\bar{J}+J_0\right]$.
Such optimized pulses will realize the Toffoli gate with least sensitivity to $J$.
In optimization problem~(\ref{eq:optimization_flattening}), while the first term in the bracket forces
the fidelity have high values in the above interval the other two terms
flatten the fidelity curve in that interval symmetrically. By decreasing the value of $\beta$ the fidelity
curve becomes more flat in the interval. Specially, for the set $\mathbf{u}_5$ ($\beta\approx0.1$)
when $|J/\bar{J}-1|\leq 0.1$ the variation in the fidelity belongs to $[0.918, 0.919]$.

The sets $\mathbf{u}_3$, $\mathbf{u}_4$ and $\mathbf{u}_5$ are depicted in Fig.~\ref{fig:fid_ver_j}
by curves marked with {\small$\blacktriangle$}, {\small$\blacklozenge$} and {\small$\bigstar$}
, respectively.

In Sec.~\ref{sec:mf_fidelity}, we analyze the fidelity multifractal behavior for the pulses in each
of the sets $\mathbf{u}_1$ to $\mathbf{u}_5$.

\section{\label{sec:multifractal}Multifractal Analysis}


Fractal dimension is an index 
that informs how detail in a pattern changes depending on the scale it is~\cite{mandelbrot1982}, which can be promptly associated to regularity. This way, fractal analysis provides a framework for characterization and modeling of irregular traces and complex shapes found in nature~\cite{mandelbrot1968fractional,voss1989random}. However, many phenomena 
that have been identified in physics and applied sciences do exhibit scaling behavior with wild regularity variations which cannot be completely characterized by a single fractal dimension, but with a entire spectrum of fractal dimensions~\cite{stanley1988multifractal,arneodo1995thermodynamics}. In face of this difficulty, the multifractal formalism was proposed as a way of characterizing such form of complexity in terms of the scaling properties of singularity measures~\cite{parisi1985fully,halsey1986fractal}. 

The multifractal formalism consists in determining a singularity spectrum $f(\alpha)$, where the singularity strength $\alpha$ accounts for the local regularity and $f(\alpha)$, the Hausdorff dimension of $\alpha$, gives a geometrical idea of the repartition of these singularities~\cite{halsey1986fractal,barral2005multifractal}. 

In general, $f(\alpha)$ is not assessed directly from data but via a scaling function, such as $\zeta(q)$, which is then connected to $f(\alpha)$ by the Legendre transform~\cite{parisi1985fully}
\begin{equation}
     \alpha = d\zeta(q)/dq, \qquad f(\alpha) = \alpha q - \zeta(q) +1,
    \label{eq:legendre_transf}
\end{equation}
where $\zeta(q)$ is the power-law exponent of a structure function of order $q$~\cite{parisi1985fully}.

A number of empirical multifractal formalisms are available in the literature. (For review and comparison of distinct methods c.f.~\cite{*turiel2006numerical,*oswiecimka2006wavelet,*huang2011arbitrary,welter2013multifractal}.) 
In this study a recently proposed formalism~\cite{welter2013multifractal}, which is briefly described in the sequel, is employed for analysis.

\subsection{EMD-DAMF}

 The EMD-based dominant amplitude multifractal formalism (EMD-DAMF) is a moment-based method. In layman terms, a dominant amplitude multifractal formalism involves initially a multiscale decomposition of the signal of interest. Then, a search for high magnitude events across different scales is run to form the set of the so-called dominant amplitude coefficients. In the EMD-DAMF, for a given time scale $k$, the structure functions $S_k(q)$ are defined as $q$-order statistical moments of a set of dominant amplitude coefficients $v_{k,\cdot}$, i.e., 
  \begin{equation}
    S_k(q) \coloneqq \langle \left( v_{k,\cdot} \right)^{q} \rangle =\frac{1}{n_k}\sum_{i=1}^{n_k} \left(v_{k,i}\right)^{q}.
      \label{eq:moment_function}
  \end{equation}


The dominant amplitude coefficients are obtained via the empirical mode decomposition (EMD)~\cite{huang1998empirical}, as explained below. 

The EMD is a data-driven procedure which decomposes a multicomponent time series $X(t)$ in a relatively small number of multiscale components called intrinsic mode functions (IMFs) and a monotonic trend: $X(t)=\sum_k c_k(t) + r(t)$. 
Each IMF can be written as $c_k(t)=a_k(t)\cos\,\varphi_k(t)$, where $a_k(t)$ is a slowly varying amplitude and $\varphi_k(t)$ is the instantaneous phase~\cite{huang1998empirical}.

One advantage of employing the EMD as a multiscale decomposition is that, thanks to its data-driven formulation, it naturally adapts to signal features and time scales. Moreover, the EMD involves computing signal envelopes so that when an IMF $c_k(t)$ is obtained, $|a_k(t)|$ is already available. Hence, searching for high magnitude events across different time scales can be accomplished by looking for the local maxima of $|a_k(t)|$.
In order to avoid arbitrary small values of amplitude, which could lead to divergence of negative moments in Eq.~(\ref{eq:moment_function}), the dominant amplitude coefficients are defined as~\cite{welter2013multifractal}
  \begin{equation}
    v_{k,i} \coloneqq \sup_{k'\le k} \left\{\max \left(\,\left| a_{k'}(t\in I_{k,i}) \right|\,\right) \right\},
    \label{eq:dam}
  \end{equation}
for $k=1,\,2,\dots$, with $i=1,\dots,\,n_k$, where $n_k$ is number of local maxima of $a_k(t)$, and $I_{k,i}$ is a time support around the $i$th maxima of $a_k(t)$.

For processes presenting scaling properties one can expect that $S_k(q) \simeq \tau_k^{\,\zeta(q)}$ for $k_{\mathrm{min}} \le k \le k_{\mathrm{max}}$, where $\tau_k$ is the mean timescale of the $k$th component. Hence, the singularity spectrum can be estimated from Eq.~(\ref{eq:legendre_transf}). The novelty of the EMD-DAMF method relies on the proper choice of the multiresolution coefficients $\{ v_{k,\cdot} \}$ which permits to estimate $\zeta(q)$ even for negative values of $q$ and, consequently, to obtain both sides of the $f(\alpha)$ spectrum~\cite{welter2013multifractal}. (See \footnote{\texttt{http://lps.lncc.br/index.php/demonstracoes/emd-damf}} for a computer program with examples of EMD-DAMF.) 

\subsection{Singularity spectrum attributes and complexity}

Considering the general complexity of engineered quantum gates and their interaction with noise, it is reasonable to expect that fidelity time fluctuations may reflect system complexity in some manner. 

\begin{figure}
\includegraphics[trim = 0mm 0mm 0mm 0mm, clip=true, width=1.0\columnwidth]{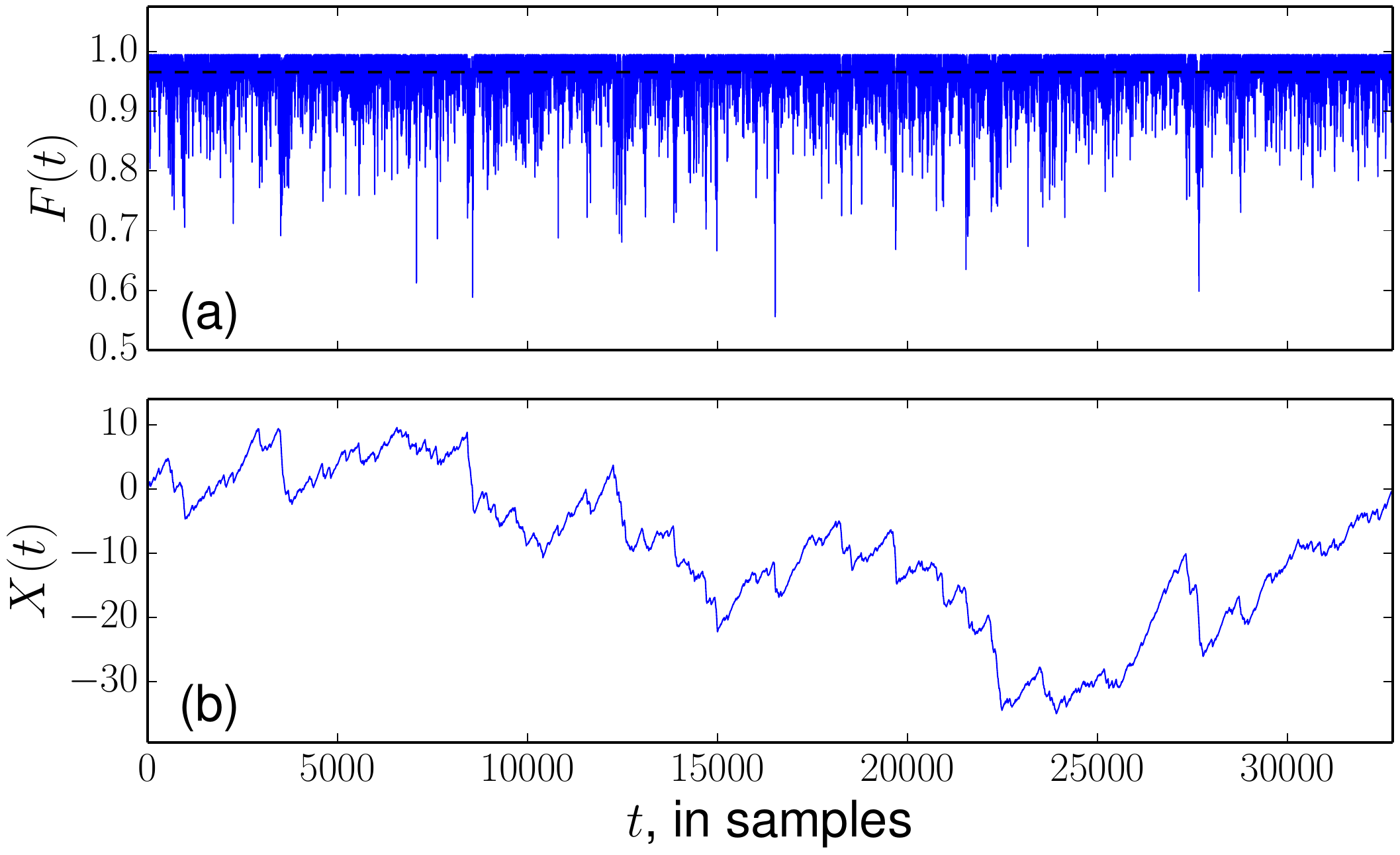}
\caption{\label{fig:mf_signals}(Color online) Fidelity time fluctuations $F(t)$ in (a) and its integrated path $X(t)=\int_{0}^{t}[F(t')-\langle F \rangle]dt'$ in (b). The dashed black line in (a) is the mean fidelity $\langle F \rangle$.}
\end{figure}

A typical realization of the fidelity $F(t)$ of the Toffoli gate exhibits an apparent random behavior, as it can be seen in Fig.~\ref{fig:mf_signals}(a). Since $F(t)$ is in general a poorly correlated signal, it is advisable to perform multifractal analysis in its integrated path, $X(t)=\int_{0}^{t}[F(t')-\langle F \rangle]dt'$, which is shown in Fig.~\ref{fig:mf_signals}(b).

\begin{figure}
\includegraphics[trim = 0mm 0mm 0mm 0mm, clip=true, width=1.0\columnwidth]{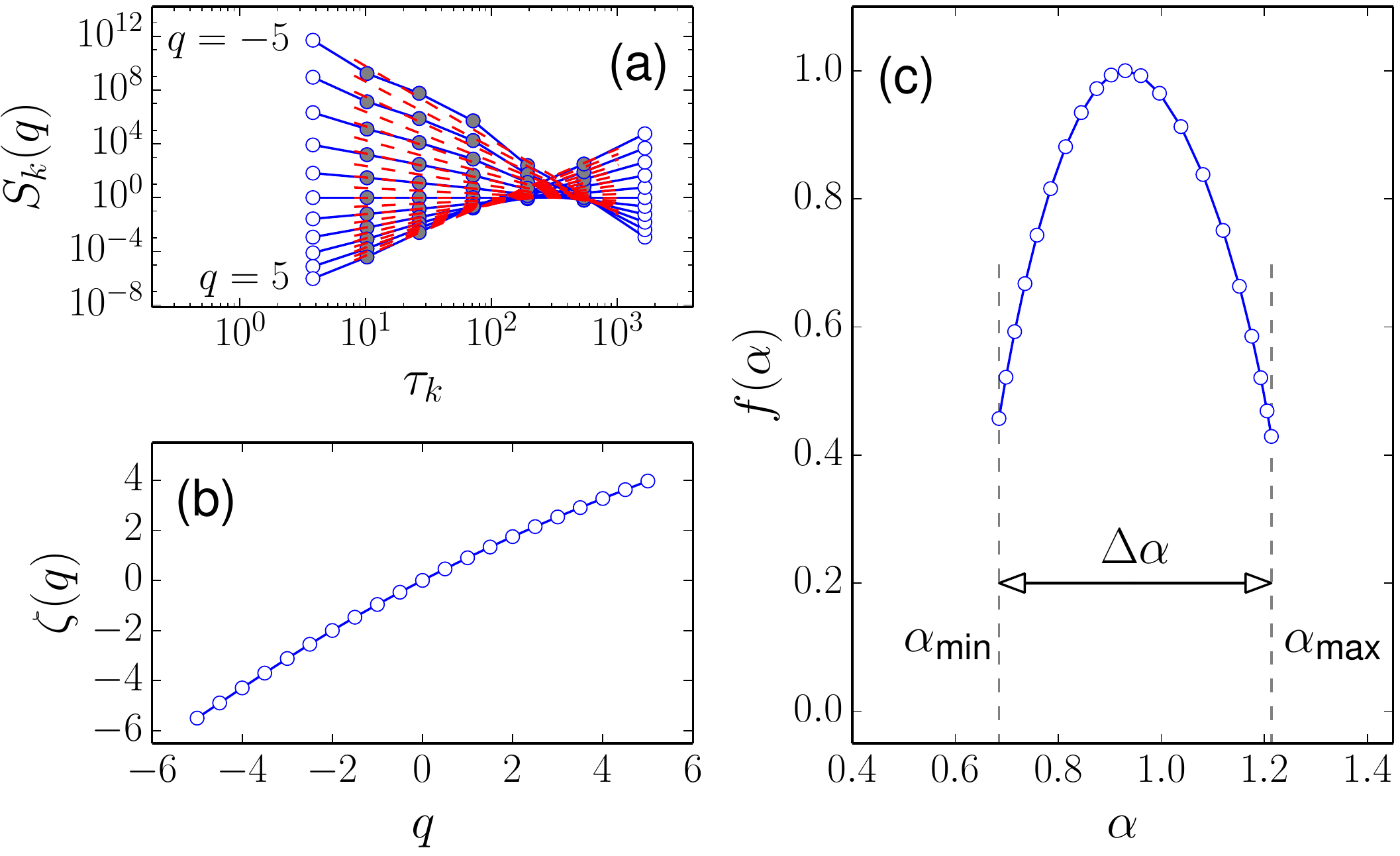}
\caption{\label{fig:mf_fidelity11}(Color online) In (a), the scaling function $S_k(q)$ obtained for the moments $q=-5,-4,\dots,5$ from the integrated fidelity path $X(t)$ showed in Fig.~\ref{fig:mf_signals}(b). The dashed red lines in (a) represent the power-law fit of $S_k(q)$ obtained with $k_{\text{min}}=2$ and $k_{\text{max}}=6$ for each value of $q$ and, in (b), their corresponding scaling exponent $\zeta(q)$. Finally, the singularity spectrum $f(\alpha)$ is obtained from $\zeta(q)$ through Eq.~(\ref{eq:legendre_transf}) and it is shown in (c).}
\end{figure}

The application of the EMD-DAMF method is exemplified in Fig.~\ref{fig:mf_fidelity11}, where one sees in panel (a): the moment function $S_k(q)$ and its scaling behavior; in panel (b): the
corresponding scaling exponents $\zeta(q)$, for $q$ between -5 and 5, in steps of 0.5; and, finally, in panel (c): its corresponding singularity spectrum $f(\alpha)$.


The value of $\alpha$ for which $f(\alpha)$ is maximum can be roughly related to the (fractal) Hausdorff dimension of the set~\cite{halsey1986fractal}, hence, it gives a measure of the apparent smoothness of the process. Small values of $\alpha$ correspond to events with irregular fluctuations, and large values correspond to smoother fluctuations. The spectrum width $\Delta\alpha =\alpha_{\text{max}}-\alpha_{\text{min}}$, on the other hand, quantifies the richness of multifractality, therefore, $\Delta \alpha$ can be regarded as a measure of complexity. 
Furthermore, an asymmetric shape of $f(\alpha)$ can be also associated to complexity, since it indicates an unbalanced contribution of singularities~\cite{halsey1986fractal,shimizu2002multifractal}.

\section{\label{sec:mf_fidelity}Multifractal Fidelity}
In Sec.~\ref{sec:toffoli}, five different sets of control pulses for implementing the
Toffoli gate were designed. Here, the \nobreak multifractal properties of the fidelity
of those gates are analyzed.

Considering the control pulses in the set $\mathbf{u}_k$, defined in Sec.~\ref{sec:toffoli}, a large
number of Toffoli gates are realized. That is done by applying the corresponding pulses in $\mathbf{u}_k$
to system Hamiltonian~(\ref{eq:XY-type}), but with a stochastic term added to the interqubit couplings in
each realization.
The couplings are supposed to obey the relation
$J_{12}=J_{23}=6J_{13}={J(t)}$, where
\begin{equation}
 \label{eq:noisy_coupling}
 J(t) = \bar{J} \left[ 1 + \epsilon (t) \right],
\end{equation}
with $\epsilon(t)$ being a sample realization of a random process with the expectation $E[\epsilon]=0$ and
the variance $E[\epsilon^2] = \nobreak \sigma^2 < \infty$. Each sample realization $\epsilon(t)$ is set as
a $1/f$ noise.
Such a noise model can be associated with an imperfect system with dynamical imperfection
in which the noise term changes at a rate $1/t_g$~\cite{moqadam2013analyzing}.

For a given standard deviation $\sigma$ we generate  $n_r=\nobreak100$ independent sequences of the
$1/f$ noise $\epsilon(t)$ each one with $2^{15}$ samples.
Using Eq.~(\ref{eq:noisy_coupling}),
$100$ different sequences $J(t)$ are then obtained. From there the corresponding
sequences of the fidelity [Eq.~(\ref{eq:fidelity})] can be calculated
\begin{equation}
 F^{(k)}(t) = F\left( J(t) ,\mathbf{u}_k \right).
\end{equation}
Now, for each fidelity sequence the multifractal width $\Delta \alpha$ is
obtained from its corresponding integrated path $X(t)$. In all cases reported below, EMD-DAMF has been used and the
scaling function $S_k(q)$ was obtained for the moments $q$ from -5 to 5, in steps of 0.5.
The process is then repeated for different values of $\sigma$ from 0.1 to 0.5, in steps of 0.01.

\begin{figure}
\includegraphics[trim = 13mm 68mm 20mm 65mm, clip=true, width=9cm]{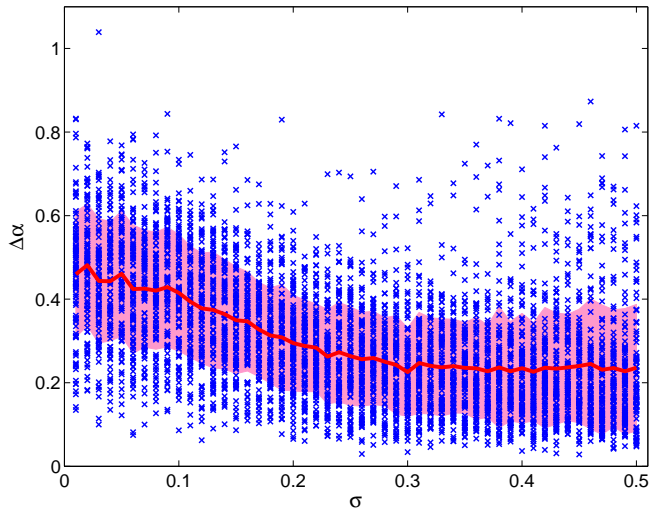}
\caption{\label{fig:delta_alpha_one_field}(Color online) The multifractal width $\Delta \alpha$
versus the standard deviation $\sigma$ calculated for 100 sample realizations of the fidelity
associated with the fields $\mathbf{u}_1$ ($\times$ points).
The red solid line shows the average behavior and the shaded area shows the
band of 1~standard~deviation around the average.}
\end{figure}

Figure~\ref{fig:delta_alpha_one_field} shows the estimated multifractal width versus standard deviation
$\sigma$, for each of the $n_r=100$ sequences of the fidelity. The fidelities correspond to the
control pulses $\mathbf{u}_1$.
For any given $\sigma$ the set of $100$ instances of $\Delta \alpha$'s has its own average and standard
deviation.
The red solid line depicts such an average and the shaded area shows the band of 1 standard deviation
around the average. The number of $\Delta \alpha$ estimates within the band for each $\sigma$ is larger
than 60. The figure shows that the average multifractality decreases when the standard deviation of the noise
increases.

Now, we calculate $F^{(k)}(t)$ for all five sets of control pulses, defined in  Sec.~\ref{sec:toffoli},
using the above method systematically. Here, for each set of control pulses, the same ensemble
of $n_r=100$ sample realizations of $J(t)$ is used.

\begin{figure}
\includegraphics[trim = 13mm 68mm 20mm 65mm, clip=true, width=9cm]{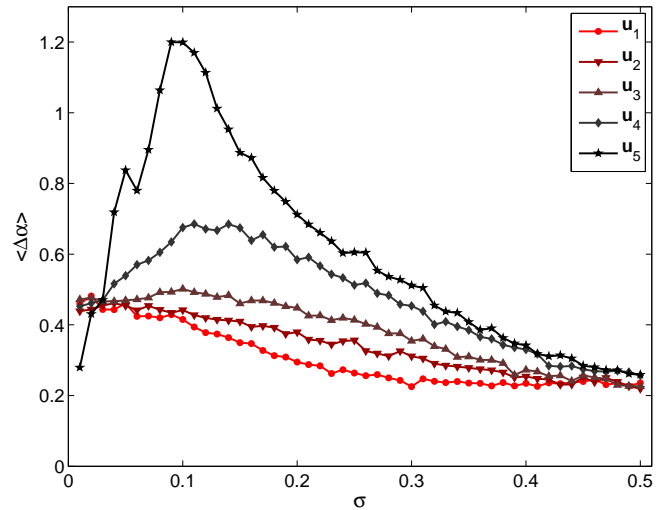}
\caption{\label{fig:delta_alpha}(Color online) The average multifractal width versus the standard
deviation $\sigma$ for five different sets of control fields
(see the text for the definitions of $\mathbf{u}_1$ to $\mathbf{u}_5$).}
\end{figure}

Figure~\ref{fig:delta_alpha} shows the average multifractal width as a function of the standard
deviation for the five sets of control pulses. The average is calculated over the $100$ estimates of
$\Delta \alpha$ for each value of the standard deviation.
As can be seen, for $\sigma=0.1$ the value of $\Delta\alpha$ increases
from $\mathbf{u}_1$ to $\mathbf{u}_5$.


Here, as regards $\mathbf{u}_3$, $\mathbf{u}_4$ and $\mathbf{u}_5$, by
reducing $\beta$ in Eq.~(\ref{eq:optimization_flattening}), respectively, we progressively flattened 
the fidelity curve in the interval
$|J/\bar{J}-1|\leq 0.1$. As a consequence, we observe an increase in the average multifractal width
around $\sigma=0.1$, from  $\mathbf{u}_3$ to $\mathbf{u}_5$. A possible explanation for the observed behavior
lies in
the more complex control effort associated with $\mathbf{u}_5$, when it tries to render $F\left(J(t),
\mathbf{u}_5\right)$ less
sensitive to variations in $J$ around the nominal values. Since $F$ is a function of $\mathbf{u}$,
the complexity
of $\mathbf{u}$ carries over to $F$.
The average multifractal widths at $\sigma=0.1$ for the sets $\mathbf{u}_1$ to $\mathbf{u}_5$
are given by 0.4152, 0.4418, 0.5018, 0.6751 and 1.1993, respectively.
Therefore, the related value for the set $\mathbf{u}_5$ increases by a factor of about
2.9 with respect to the the set $\mathbf{u}_1$. The control pulses in the set $\mathbf{u}_5$
still implement the Toffoli gate with a fidelity around $92\%$ which is
acceptable~\cite{stojanovic2012quantum}.

Figure~\ref{fig:delta_alpha} shows that the average multifractality decreases when the noise
strength in $J$ increases. It suggests that the multifractality is destroyed when the noise strength
increases~\cite{Dubertrand2014two}.
Moreover, the curves apparently converge to the same value for sufficiently large $\sigma$.
A possible explanation for such behaviors of multifractality for large $\sigma$ would be the following.
When the noise variance in $J$ increases well beyond the tolerance $J_0$ set in Eq.~(\ref{eq:optimization_flattening}),
none of the five control pulses, regardless of their complexity is capable of guaranteeing a well-behaved fidelity in
terms of $J$. Again, since $F$ is a function of both $J$ and $\mathbf{u}$, it is likely the variance of $J$ will dominate
over the complexity of $\mathbf{u}$, resulting in a low complexity in $F$.


If instead of the $1/f$ noise, white Gaussian noise was used, similar results to Fig.~\ref{fig:delta_alpha}
would be attained but generally with lower values of $\Delta \alpha$. However, even in this case,
there is still considerable amount of multifractality for the set $\mathbf{u}_5$. The curve corresponding
to the set $\mathbf{u}_5$ reaches the maximum $\langle \Delta \alpha \rangle = 0.9936$
at $\sigma=0.07$ (not shown).
As before, the average multifractality curves decrease when $\sigma$
increases and apparently converge to the same value for sufficiently large $\sigma$. The limiting
value in this case, however, is lower than before.

The multifractality observed here seems to be of a different origin than in Anderson transitions
corresponding to localization critical phenomena~\cite{evers2008anderson}.
The system can tolerate $10\%$ deviation in the value of the couplings and the gate can be still
realized with relatively high fidelity. In the fidelity of the Toffoli gate, as described above, the
multifractality reflects the complexity of the control fields implementing the gate.

\section{\label{sec:conclusion}Summary and Discussion}
In this paper, we have analyzed the multifractal width of the fidelity of five different realizations
of the Toffoli gate.
We considered a system of three coupled qubits described by the Heisenberg $XY$ Hamiltonian.
The Toffoli gate can be realized in such a system by applying
Zeeman-like control fields. When defining the fidelity functional, we considered five different optimization
problem and found the corresponding global optimized control fields. All the attained sets of control
pulses can implement the Toffoli gate with the fidelity above $90\%$, in the noiseless case.
However, each set has a different
sensitivity to variations in the interqubit couplings.

We realized the Toffoli gate for a large number of times by applying each set of control pulses to
an imperfect system whose couplings affected by additive $1/f$ noise. It was supposed the couplings
changed in each realization but remained fixed during the gate implementation.
For each of the 100 sample realizations of the fidelity, and for a given standard deviation associated
with the couplings, estimates of the  multifractal width of the fidelity sequences were obtained via
the EMD-DAMF formalism.
We found that flattening the fidelity functional with respect to $J$
around its maximum implies an increase in the estimate of its average multifractal width.
The flatter $F$ is around $J=\bar{J}$, the larger the measured average multifractality
of $F$, for $J$ contaminated with $1/f$ noise with standard deviation $\sigma$.
We observed that for noise standard deviations above 0.1, the average multifractality tends
to decrease with $\sigma$ and apparently converges to a fixed value for sufficient high $\sigma$.  

The multifractality behavior observed here is a result of the complexity in the
implementation of the gate. The effects of the number of control pulses may be
analyzed to see if there is a critical number for the control pulses below which
no multifractality can be observed.
The approach that is given in Sec.~\ref{sec:toffoli} for implementing the Toffoli
gate is a standard way in quantum control theory in engineering quantum gates.
Therefore, the multifractality behavior of the fidelity observed here is also
expected in other engineered gates. Specially, it is interesting to check the
multifractality for the $CNOT$ gate which is less complex.

In our analysis, the measure of multifractality was taken over the gate fidelity.
The quantum fidelity can be analyzed as well, by calculating the wave function overlap
in each gate realization. Fixing an initial state, the quantum fidelity is given by the
overlap between the two final states obtained by the evolution of the ideal and the
imperfect systems. In this way, it is possible to check whether the complexity in the
control fields introduces multifractality to the wave function or not.
Such analysis may be addressed in a future investigation.

\section*{Acknowledgments}
JKM and GSW acknowledge financial support from Brazilian National Council for Scientific and
Technological Development (CNPq). JKM acknowledges grant PCI-DB 302866/2014-0 from the same agency.

\bibliography{toffoli_mf_1,toffoli_mf_2,toffoli_mf_3}

\end{document}